\documentclass[aip,twocolumn,superscriptaddress]{revtex4}
\usepackage{graphicx}
\usepackage{amsfonts,amsmath,amssymb,amsthm,amscd}
\usepackage[T2A]{fontenc}
\usepackage{color}
\usepackage{dcolumn}
\usepackage{bm}
\usepackage{array}
\usepackage{float}

\begin{document}


\title{Ion Acceleration in Laser Generated Mega Tesla Magnetic Vortex}

\author{Jaehong Park}
\email{jaehongpark@lbl.gov} 
\affiliation{Lawrence Berkeley National Laboratory, Berkeley, California 94720, USA}

\author{Stepan S. Bulanov}
\affiliation{Lawrence Berkeley National Laboratory, Berkeley, California 94720, USA}

\author{Jianhui Bin}
\affiliation{Lawrence Berkeley National Laboratory, Berkeley, California 94720, USA}

\author{Qing Ji}
\affiliation{Lawrence Berkeley National Laboratory, Berkeley, California 94720, USA}

\author{Sven Steinke}
\affiliation{Lawrence Berkeley National Laboratory, Berkeley, California 94720, USA}

\author{Jean-Luc Vay}
\affiliation{Lawrence Berkeley National Laboratory, Berkeley, California 94720, USA}

\author{Cameron G.R. Geddes}
\affiliation{Lawrence Berkeley National Laboratory, Berkeley, California 94720, USA}

\author{Carl B. Schroeder}
\affiliation{Lawrence Berkeley National Laboratory, Berkeley, California 94720, USA}

\author{Wim P. Leemans}
\affiliation{Lawrence Berkeley National Laboratory, Berkeley, California 94720, USA}

\author{Thomas Schenkel}
\affiliation{Lawrence Berkeley National Laboratory, Berkeley, California 94720, USA}

\author{Eric Esarey}
\affiliation{Lawrence Berkeley National Laboratory, Berkeley, California 94720, USA}



\begin{abstract}
Magnetic Vortex Acceleration (MVA) from near critical density targets is one of the promising schemes of laser-driven ion acceleration. 
3D particle-in-cell simulations are used to explore a more extensive laser-target parameter space than previously reported on in the literature as well as to study the laser pulse coupling to the target, 
the structure of the fields, and the properties of the accelerated ion beam in the MVA scheme.
The efficiency of acceleration depends on the coupling of the laser energy to the self-generated channel in the target. 
The accelerated proton beams demonstrate high level of collimation with achromatic angular divergence, and carry a significant amount of charge. For PW-class lasers, this acceleration regime provides favorable scaling of maximum ion energy with laser power for optimized interaction parameters. 
The mega Tesla-level magnetic fields generated by the laser-driven co-axial plasma structure in the target are prerequisite for accelerating protons to the energy of several hundred MeV.
\end{abstract}


\maketitle

\section{Introduction}

Laser-based ion acceleration (see \cite{MTB_review,Daido_review,Macchi_review,Bulanov_review} and references cited therein) has received considerable attention over the last two decades for the potential applications to diverse research areas: fundamental particle physics, inertial confinement fusion, warm-dense matter, medical therapy, etc. It is expected that with the fast development of multi-PW laser facilities \cite{BELLA,ELI-NP,ELI-BL,CORELS,XCELS,VULCAN,Danson_2015} laser ion acceleration will be able to generate ion beams with energies in excess of 100 MeV, required by many applications. Up to now laser systems were only able to achieve the acceleration of ions with energies approaching 100 MeV \cite{RPA_93 MeV,TNSA_85 MeV,TNSA_RPA_94 MeV}. While most of the experimental results were obtained in the Target Normal Sheath Acceleration (TNSA) regime \cite{snavely2000,maksimchuck2000,Wilks_2001,Fuchs_2006},
higher ion energies are expected to be generated by employing advanced regimes of laser ion acceleration, as it was demonstrated in Refs. \cite{RPA_93 MeV, TNSA_RPA_94 MeV}. These regimes include, to name a few, Radiation Pressure Acceleration (RPA) \cite{RPA,henig2009,bin2015}, Shock Acceleration (SA) \cite{SWA}, Relativistic Transparency (RIT) \cite{RIT}, and Magnetic Vortex Acceleration \cite{MVA}. Analytical and computer simulation estimates show that a PW or several PW laser system may be able to generate ions with energies ranging from several hundred MeV to GeV per nucleon (see \cite{SSB_PoP} and references cited therein). We note that NCD targets as well as composite targets with NCD parts attracted a lot of attention recently not only to be used for ion acceleration \cite{bulanov.prl.2015,bin2015}, but also for brilliant gamma-ray and electron-positron pair production \cite{brady.ppcf.2013, zhu.natcomm.2016, liu.pop.2018, gu.commphys.2018}. All these results rely on the physics of intense laser pulse interaction with NCD plasma, the basics of which are best illustrated by the MVA. 

In this paper we study the MVA regime for a PW-class laser system. This regime uses near-critical density (NCD) slabs as targets, in contrast to thin micron or sub-micron solid density foils used in other regimes. 
Experimental studies in such targets have reported maximum ion energy of several tens of MeV per nucleon at sub-PW laser systems \cite{willingale2006,fukuda2009,willingale2011,helle2016} and previous 2D/3D computer simulation studies showed that the maximum ion energy can reach GeV level with PW-class laser systems \cite{bulanov2010,  nakamura2010, bulanov2015, sharma2016, sharma2018}. 

In the MVA scheme, an intense laser beam can penetrate the NCD target and expels the electrons by the ponderomotive force. It thereby creates a low density channel in the electron plasma component along the laser propagation axis while the ions remain at rest due to their larger mass during a short amount of time. The radius of the channel can be determined from balancing the energy gain of an electron in a laser field and in the field of an ion column \cite{, bulanov2015}:
\begin{equation}
R_\text{ch}={\lambda\over\pi}\left({n_\text{cr}\over n_e}\right)^{1/3}\left({2\over K}{P\over P_c}\right)^{1/6},
\label{rcheq}
\end{equation}
where $n_e$ is the electron density, $n_\text{cr}= m_e \omega^2/4\pi e^2$ is the critical density, $e$ and $m_e$ are the charge and mass of an electron respectively, $\omega$ is the laser angular frequency, $K=1/13.5$ is the geometrical factor, $P$ is the laser power, $P_c=2m_e^2c^5/e^2=17$GW is a characteristic power for relativistic self-focusing \cite{sun1987}, and $c$ is the speed of light in vacuum.

As the laser propagates in the self-generated channel, it accelerates electrons in its wake. These electrons form a thin filament along the central axis, carrying strong electric current, which is due to the plasma lensing effect \cite{chen.ieee.1987}, {\it i.e.}, to the electron flow pinching as it propagates through the ion background \cite{bulanov2015}. Thus, a co-axial plasma structure is formed with the current flowing along the axis and the return current flowing in the channel wall, resulting in a strong azimuthal magnetic field confined inside the channel (we note that a similar approach to generating strong azimuthal magnetic fields in plasma was reported in Refs. \cite{stark2016,jansen2018,arefiev2018}, where a long laser pulse was interacting with a pre-filled channel). When the laser, followed by this pinched current, exits the target from the back, the magnetic fields begins to expand in the transverse direction. In doing so, the field displaces the electron component of the plasma with respect to the ion one, and, as a result, both strong longitudinal and transverse electric fields are generated, which accelerate and collimate ions in the form of a well defined beam with achromatic divergence. These accelerated ions mainly originate from the same filamentary structure, since the electron current pre-accelerates a number of ions as it propagates through the ion channel. 

The maximum achievable ion energy in the MVA scheme is determined by several parameters such as target density, target length, laser power, laser focal spot size, etc. The optimum condition to maximize the proton energy for a given laser power can be obtained based on the waveguide model, where the electromagnetic energy is perfectly confined inside the self-generated channel  \cite{bulanov2010}. The optimum condition for acceleration is given by the following relation, basically obtained by equating the laser energy to
the total energy the electrons inside the channel can acquire after interacting with the laser \cite{bulanov2010}:
\begin{equation}
{n_e\over n_\text{cr}}=2^{1/2}K\sqrt{{P\over P_c}}\left({L_p\over L_\text{ch}}\right)^{3/2},
\label{opteq}
\end{equation}
where $L_p=c\tau$ is the laser pulse length, $\tau$ is the laser pulse duration, and $L_\text{ch}$ is the target length.

Most previous studies of the MVA scheme have been done through 2D Particle-in-Cell (PIC) simulations that successfully qualitatively explained how the mechanism works, though in was understood that the magnetic vortex is a 3D structure  \cite{MVA, bulanov2010, nakamura2010}. So, in order to get quantitatively accurate results on the MVA scheme, 3D PIC simulations are required. Recent 3D simulations of the MVA scheme \cite{helle2016,sharma2016,sharma2018} explored ion acceleration for different laser powers and  polarization (linear and circular cases), but left the study of the coupling and field structure out.

In this paper, we explore the MVA scheme using 3D PIC simulations in a more extensive parameter space: we vary the laser power, the laser focal spot size, target density, and pulse duration. 
Here, we focus on the study of laser pulse coupling to the target, the structure of the fields in the target, as the laser propagates through it, the scaling of the maximum ion energy with laser parameters, such as power and duration, as well as on the properties of the accelerated ion beam, which is of great importance for applications and beam transport.
 
We show that the intense laser interaction with an NCD target creates a co-axial plasma current structure, which generates a localized mega Tesla-level magnetic field. 
The converging electric field behind the rear-surface of the target makes a contribution to the collimation of the ion beam and therefore the accelerated ions reveal achromatic divergence in the angular distribution. 
Moreover a favorable scaling of maximum ion energy is revealed when two conditions are satisfied: (i) the laser focal spot size matches the radius of the self-generated channel in Eq.(\ref{rcheq}) and (ii) the target density and the length are determined by the optimum condition in Eq.(\ref{opteq}).

The rest of the paper is organized as follows. The simulation setup and the parameter space are described in Section II. The simulation results are in Section III and IV. The summary and conclusion are in Section V.

\section{3D PIC simulation setup}

\begin{table}[t]
\begin{tabular}{l*{7}{c}r}
\hline
$\text{Run}$ & $a_0$  & $\tau$ (fs) & $n_e/n_\text{crit}$ & $w_0$($\mu$m) & P(PW) & $E_L(J)$ & $E_\text{max}$(MeV)& \\
\hline
 I-1 & 161 & 27 & 4.52 & 1.49 & 1.96 & 43.9 & 466 \\
 I-2 & 118 & 27 & 3.32 & 1.49 & 1.05 & 23.6 & 283\\
 I-3 & 91  & 27 & 2.56 & 1.49 & 0.63 & 14.0 & 182\\
 I-4 & 70  & 27 & 1.96 & 1.49 & 0.37 & 8.3 & 132\\
 I-5 & 54  & 27 & 1.51 & 1.49 & 0.22 & 4.9 & 83\\
 \hline
 II-1 & 120 & 27 & 4.52 & 2.0 & 1.96 & 43.9 & 360\\
 II-2 & 88  & 27 & 3.32 & 2.0 & 1.05 & 23.6 & 196\\
 II-3 & 68  & 27 & 2.56 & 2.0 & 0.63 & 14.0 & 153\\
 II-4 & 52  & 27 & 1.96 & 2.0 & 0.37 & 8.3 & 108\\
 \hline
 III-1 & 80  & 27 & 4.52 & 3.0 & 1.96 & 43.9 & 138\\
 III-2 & 58  & 27 & 3.32 & 3.0 & 1.05 & 23.6 & 107\\
 III-3 & 45  & 27 & 2.56 & 3.0 & 0.63 & 14.0& 85\\
 III-4 & 35  & 27 & 1.96 & 3.0 & 0.37 & 8.3 & 50\\
 \hline
 IV-1 & 114 & 13.5 & 1.60 & 2.1 & 1.96 & 21.9 & 227\\
 IV-2 & 84  & 13.5 & 1.17 & 2.1 & 1.05 & 11.8 & 140\\
 IV-3 & 65  & 13.5 & 0.91 & 2.1 & 0.63 & 7.0 & 89\\
 IV-4 & 49  & 13.5 & 0.69 & 2.1 & 0.37 & 4.2 & 55\\
\hline
 V-1 & 161 & 54 & 4.52 & 1.49 & 1.96 & 87.8 & 412 \\
 V-2 & 118 & 54 & 3.32 & 1.49 & 1.05 & 47.2 & 275\\
 V-3 & 91  & 54 & 2.56 & 1.49 & 0.63 & 28.0 & 196\\
 V-4 & 70  & 54 & 1.96 & 1.49 & 0.37 & 16.6 & 131\\
 \hline
\end{tabular}
\caption{Initial parameters of 3D simulations organized into five groups;
$a_0$: dimensionless vector potential, $\tau$: laser pulse duration, $n_e$: electron density, $w_0$: laser waist,
$P$: laser power, $E_L$: laser energy, and $E_\text{max}$: maximum ion kinetic energy.
Each group has a different laser spot size $w_0$. The laser power varies from $0.22$ to $1.96$ PW. The electron density is chosen by the optimum condition in Eq.(\ref{opteq}), except in group V. The spot sizes from groups I, IV, and V match the channel radius in Eq.(\ref{rcheq}).}
\label{setuptable}
\end{table}

\begin{figure*}
\begin{center}
\includegraphics[scale=0.5]{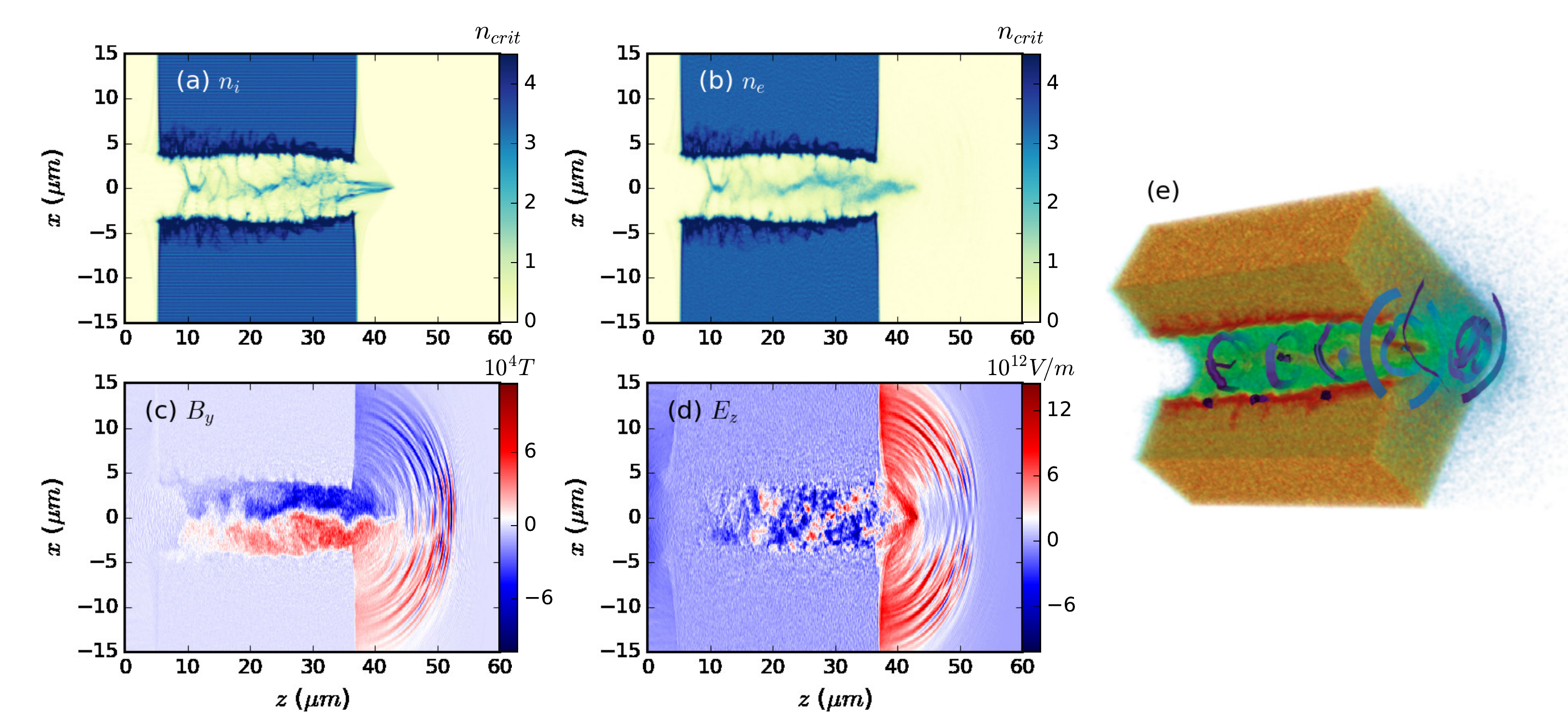}
\caption{Simulation result (Run I-2 in Table \ref{setuptable})
around the acceleration stage at $t=226$ fs:
(a) ion density $n_i/n_\text{crit}$, (b) electron density $n_e/n_\text{crit}$, (c) $B_y$ field, and (d) $E_z$ field, in the $x-z$ slice at $y=0$.
(e) 3D image of the ion density distribution;  
a strong magnetic field (ribbon) inside the low-density channel is generated.}
\label{fig:fields}
\end{center}
\end{figure*}

We use the 3D relativistic full particle-in-cell (PIC) code WarpX \cite{vay2018}.
The target is an NCD-hydrogen plasma with $n_e=0.69-4.52$ $n_\text{crit}$, with longitudinal thickness of $L_\text{ch}=32\mu$m. The target density is uniform in the range of $5\mu\text{m}<z<37\mu\text{m}$ and is zero elsewhere.

The laser pulse has both transverse and longitudinal Gaussian profile and propagates along the $z$-axis. It is tightly focused at the target front surface, $z=5\mu$m, with the focal spot size (laser waist), $w_0=1.488-3.0\mu$m (half-width at $1/e^2$ of the intensity peak). The laser wavelength is $\lambda=0.8\mu$m. A virtual laser antenna is used to inject the laser and is located within the simulation domain at $z=1\mu$m. The electric field is linearly polarized along the $x$-axis. The laser intensity is $I=2.13- 8.54\times 10^{22}\text{W}/\text{cm}^2$  and the corresponding dimensionless vector potential is $a_0\equiv eE_0/m_e\omega c=54-161$. The laser pulse duration is chosen to be $\tau=13.5$, $27$, and $54$ fs (defined as the full width at $1/e$ of the amplitude of the electric field). The laser power is $P=0.22-1.96$PW and the total laser energy is $E_L=4.9-87.8$J.

Table \ref{setuptable} shows parameter sets of our 3D simulations organized into five groups. Each group has a different focal spot size from $w_0=1.49$ to $3\mu$m. The laser pulse duration is $\tau=27$ fs in groups I--III, $13.5$ fs in group IV, and $54$fs in group V. The target densities in each group are chosen by the optimum condition in Eq.(\ref{opteq}) except in group V. Group V has the same parameters as group I except for the pulse duration. The laser spot sizes from group I, IV, and V  match the channel radius in Eq.(\ref{rcheq}), that is, $w_0=R_\text{ch}$, while groups II and III have larger focal spot sizes, $w_0>R_\text{ch}$. The maximum ion energy is listed in the table and will be discussed in the next section.

The simulation domain size is $(L_x,L_y,L_z)=(60\mu\text{m},60\mu\text{m},80\mu\text{m})$ and the number of cells is $(N_x,N_y,N_z)=(512,512,2400)$. The cell sizes are $dz=0.0417\lambda$ ($=0.3-0.55 c/\omega_\text{pe}$) and $dx=dy=0.144\lambda$ ($=1-2 c/\omega_\text{pe}$), where $\lambda$ is the laser wavelength, $\omega_\text{pe}=({n_e e^2/\epsilon m_e})^{1/2}$ is the electron plasma frequency.

The boundary conditions are periodic along the transverse directions and open along the longitudinal direction ($z$-axis) with a perfectly matched layer (PML) that absorbs outgoing waves very efficiently \cite{shapoval18}. The simulation runs until $400$ fs ($4000$ time iterations) at which the ion energy is fully saturated.
We used $1$ particle/cell/species and the simulation results converged at 
higher resolutions up to $N_x = 2400$ and $N_z=4800$.

Throughout this paper, we largely discuss the simulation result of Run I-2 in Table \ref{setuptable}, otherwise we specify the simulation parameters.

\begin{figure*}[t]
\begin{center}
\includegraphics[scale=0.45]{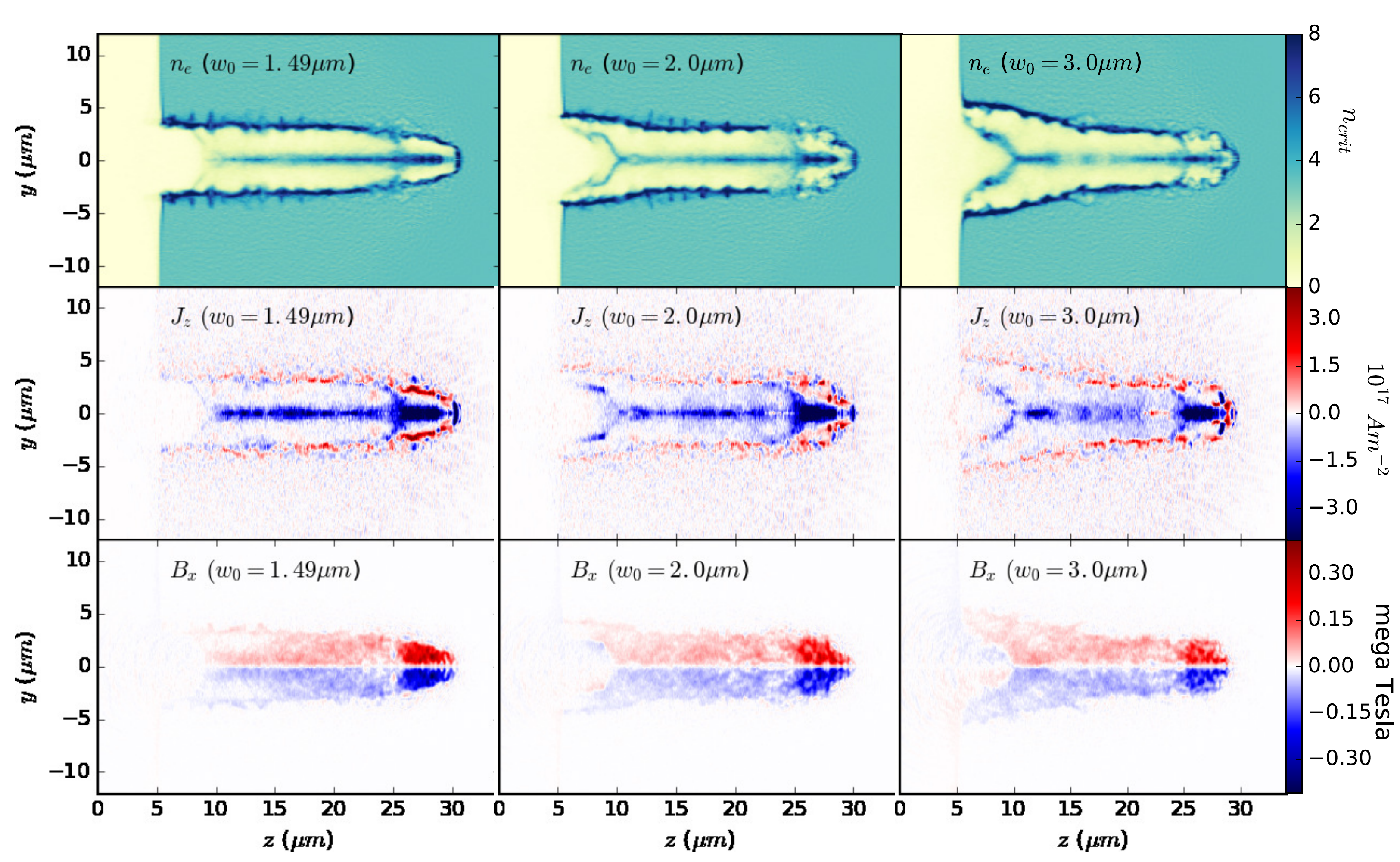}
\caption{(from top to bottom) 2D slice $y-z$ cut of
the electron density $(n_e)$, the current density $J_z$, and the magnetic field $B_x$ for different laser focal spot sizes, $w_0=1.49\mu$m (left: Run I-2), $2.0\mu$m (middle: Run II-2), and $3.0\mu$m (right: Run III-2), before the laser pulse exits the target ($t=144$fs).
In the left column panel, the laser spot size matches the leading channel radius, $w_0=R_\text{ch}$
while the other runs have $w_0>R_\text{ch}$.}
\label{fig:focal}
\end{center}
\end{figure*}

\section{Laser pulse propagation in NCD plasma}

As was mentioned above, the laser pulse makes a channel both in the electron and ion components of the plasma, as it propagates through the target, see Fig \ref{fig:fields}(a-b), where the results of Run-I for $n_e$ and $n_i$ distributions are shown at $t=226$ fs. Here $a_0=118$, $\tau=27$ fs, and $P=1.05$ PW. The propagation of the laser inside this channel is accompanied by the generation of a strong azimuthal magnetic field (Fig. \ref{fig:fields}c) and longitudinal electric field, as the laser exits the channel (Fig. \ref{fig:fields}d). Note that inside the channel, there is a pinched filamentary structure along the central axis at $x=y=0$. The filament is not perfectly straight along the central axis but wiggles along the $x$-axis as the electrons oscillate with the laser field. The filament carries the electric current toward the $-z$ direction, dominated by faster electrons, which induces the azimuthal magnetic field  [Fig.\ref{fig:fields}(c)]. As the magnetic field exits the target, it displaces the surface electrons and a strong electric field $E\sim 10-60$TV/m is induced over the distance of $\sim10\mu$m behind the rear surface of the target [Fig.\ref{fig:fields}(d)].

In order to visualize the structure of the current and magnetic field in the channel, we show in Fig. \ref{fig:fields}(e) a 3D image of the ion density distribution. The magnetic field (blue ribbon) is circulating around the filamentary structure along the central axis inside the low-density channel. The cloud near the surface of the target represents the accelerated ions.

Now, we compare the formation of the channel for different laser focal sizes.
Figure \ref{fig:focal} shows the electron density distribution (top),
the current density $J_z$ (center), and the magnetic field strength $B_x$ (bottom)
before the laser pulse exits the target ($t=144$ fs),
for different laser focal spot sizes, $w_0=1.49\mu$m (left: Run I-2), $2.0\mu$m (middle: Run II-2), and $3.0\mu$m (right: Run III-2). 
These variables are plotted in the $y-z$ slice plane to address the azimuthal component of the
magnetic field $B_\phi$, which is $B_x$ in the $y-z$ plane, distinguished from the laser field. 
Here, the laser power and the energy are fixed as $P=1.05$PW and $E_L=23.6J$ in the three runs.
Only the left column panel shows that the laser spot size matches the channel radius in Eq.(\ref{rcheq}), $w_0=R_\text{ch}$,
while the other runs have $w_0>R_\text{ch}$. The radius of the trailing channel expands as time goes by.
As the laser spot size becomes larger, 
the laser pulse is dispersed and loses the energy sideways as seen in the right column panel.
As a result, the current density of the filament along the central axis and the induced magnetic field
become weaker for $w_0>R_\text{ch}$.
The ion energy will be lowered for the larger spot sizes.
Note that the strongest magnetic field is localized inside the leading channel and
reaches around $0.4$MT (the left column panel),
while the magnetic field in the trailing channel is reduced by one order of magnitude.
As for $w_0<R_\text{ch}$ (not shown in the figure), the laser pulse would develop 
filamentation which prevents efficient channel generation as discussed in \cite{bulanov2015}.

\begin{figure*}
\begin{center}
\includegraphics[scale=0.5]{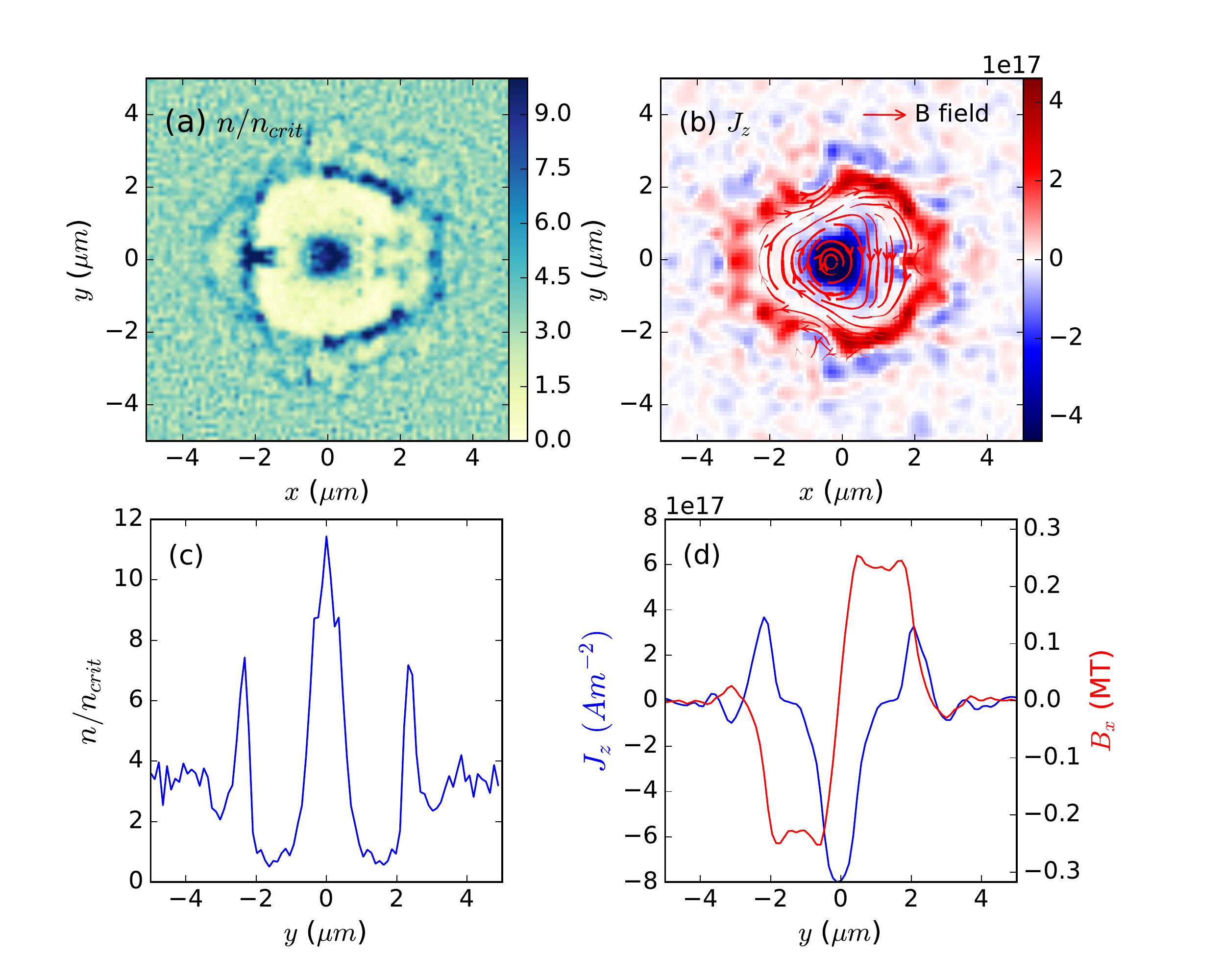}
\caption{Channel structure (Run I-2) at $t=144$fs:
(a) electron density $n_e/n_\text{crit}$, (b) electric current density $J_z$ with
the $B$ field lines (red)
in the $x-y$ slice cut on $z=28\mu$m,
(c) 1D electron density profile,
and (d) 1D electric current density $J_z$ (blue) and $B_x$ field (red) profiles
along the $y$-axis at $x=0$.}
\label{fig:channel}
\end{center}
\end{figure*}

Figure \ref{fig:channel} shows the channel structure in more detail in the $x-y$ slice cut on $z=28\mu$m at $t=144$fs  (Fig. \ref{fig:focal} left column). The channel is surrounded by thin dense walls about twice higher than the background density [Fig. \ref{fig:channel}(a and c)] and the high density bump at $x=y=0$ is the plasma pinch which carries the electric current.

Figure \ref{fig:channel}(b) shows co-axial structure of the current density $J_z$. The peak value of the current density is $J_z=8\times 10^{17}$ $I/m^2$ along the $-z$ direction and is compensated by the return current flowing in the walls of the channel, thereby perfectly screening the magnetic field outside the channel (see also Fig. \ref{fig:channel}(d)). Inside the channel, the magnetic field lines (red) are circulating around the plasma pinch with a peak value of $B\sim 0.25$ mega Tesla as seen in Fig. \ref{fig:channel}(b and d).

The intense laser interaction with an NCD target creates a co-axial plasma current structure, which generates a localized mega Tesla-level magnetic field in the leading channel. A strong magnetic field is obtained
when the laser focal size matches the channel radius, which affects ion acceleration as we will see in the next section.

\section{Ion Acceleration}

\begin{figure*}
\includegraphics[scale=0.45]{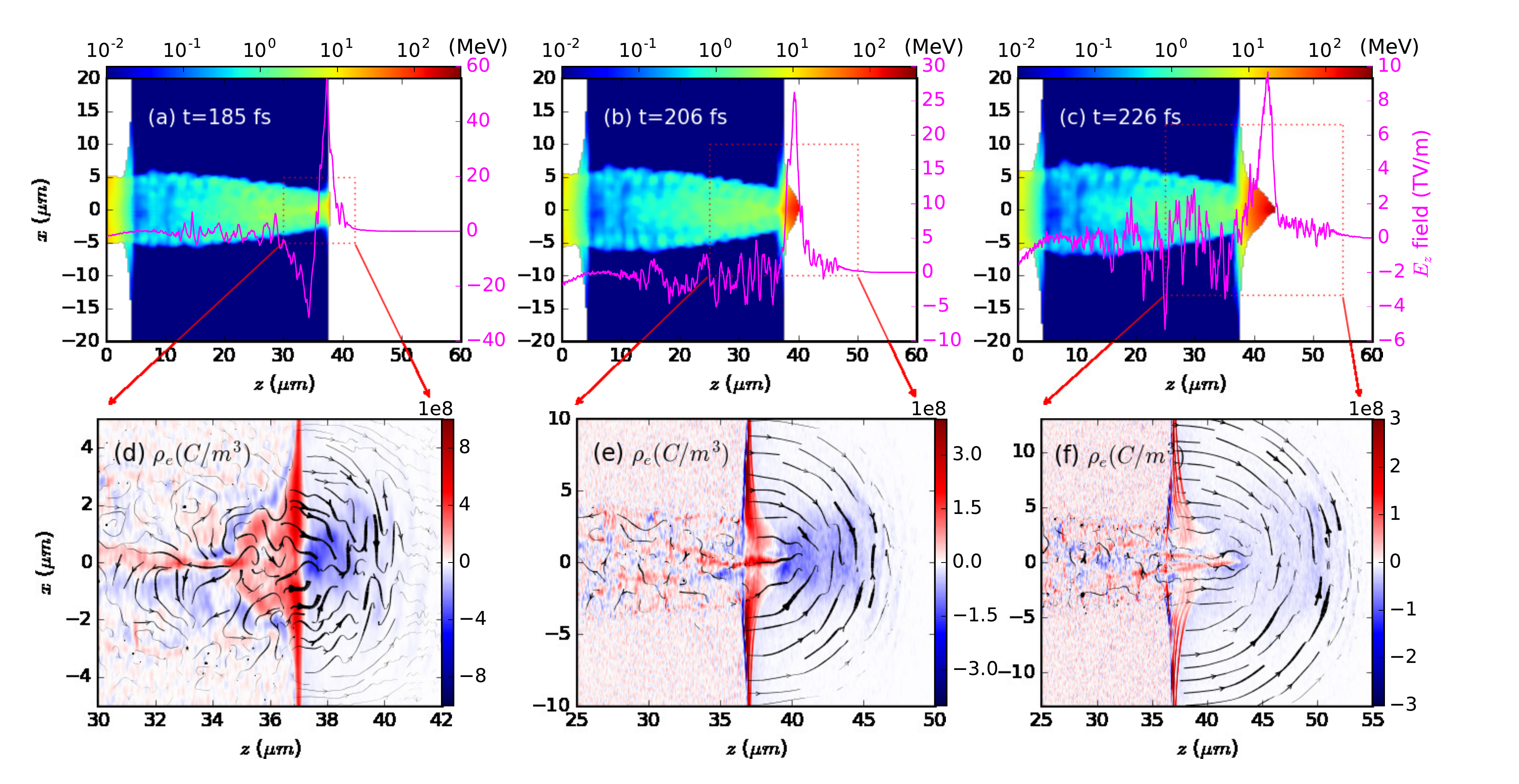}
\caption{Three consecutive time steps (Run I-2) during the acceleration stage $t=185$, $206$, and $226$ fs:
(a-c) ion energy distribution in the $x-z$ plane ($y=0$) (logarithmic color scales represent the energy in MeV) 
and the $E_z$ field along the central axis $x=y=0$ (magenta).
(d-f) charge density $\rho_e=e(n_i-n_e)$ and the electric field lines (black) $\bold{E}=E_x\hat{x}+E_z\hat{z}$ in the $x-z$ plane ($y=0$).}
\label{fig:efield}
\end{figure*}

\begin{figure*}
\begin{center}
\includegraphics[scale=0.55]{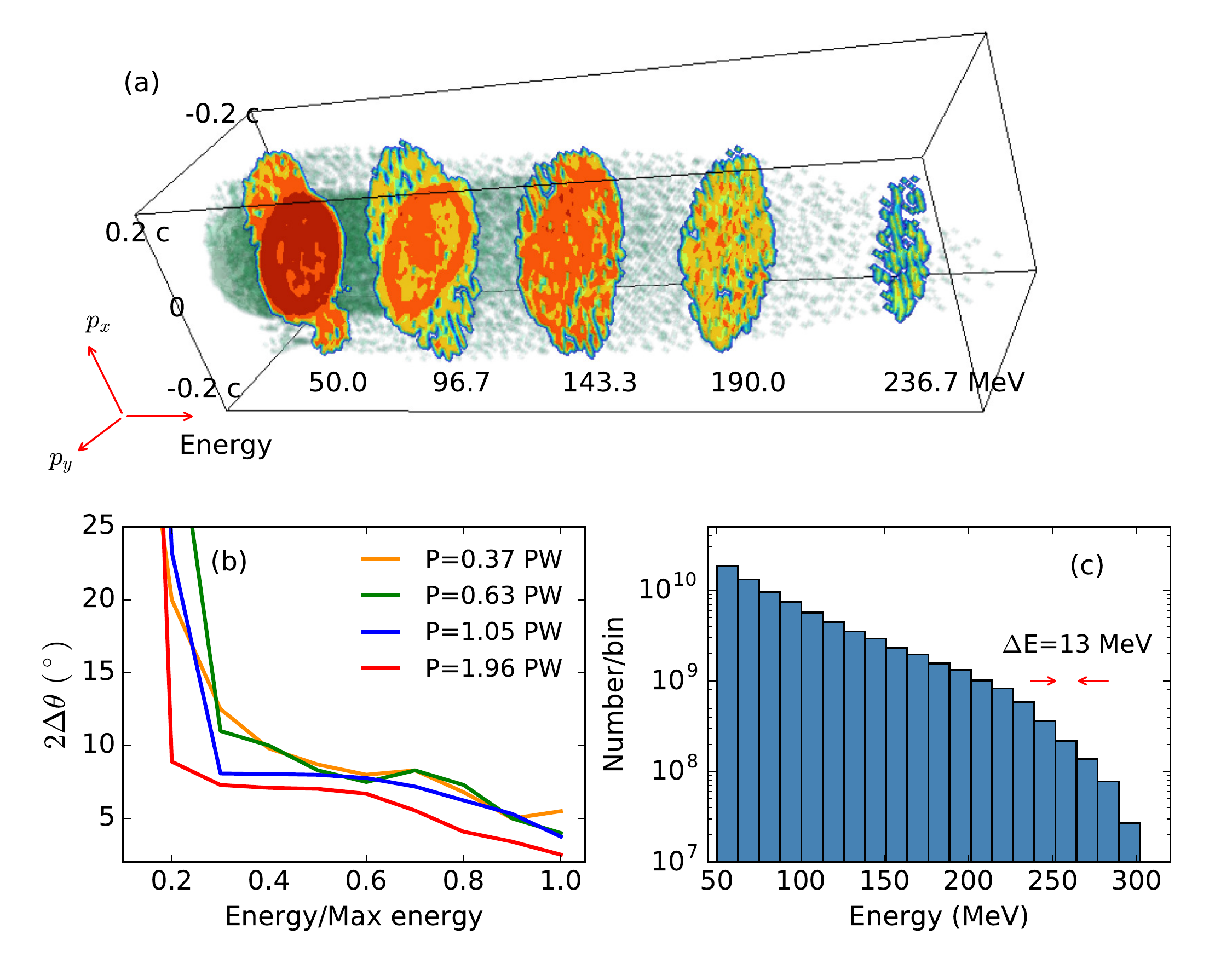}
\caption{(a) Transverse momentum $p_x-p_y$ distribution of the ions 
at several energy levels centered at $E=50,96,143,190$, and $236$ MeV (Run I-2).
(b) Divergence vs. normalized energy in the ion energy-angular distribution, for different laser powers (Run I-1 to I-4).
The energy is normalized by the maximum energy in each run.
(c) The number of the accelerated ions in each energy bin with the bin size $\Delta E=13$MeV  (Run I-2).}
\label{fig:energy}
\end{center}
\end{figure*}

In this section, we explore the properties of the accelerated ion beam for several different parameters.
Figure \ref{fig:efield} shows the ion energy distribution (top) and the charge density (bottom) at three consecutive time steps during the acceleration stage $t=185-226$ fs in the $x-z$ plane ($y=0$). In Fig. \ref{fig:efield}(a-c), the ions pinched by the electrons in a thin filament are accelerated at the rear-edge of the target (the energy is in logarithmic color-scales). The magenta line is the longitudinal electric field $E_z$ along the central axis. During this stage, the $E_z$ field strength decreases from $60$ to $10$ TV/m and the ion energy increases from $30$ to $220$ MeV. In Fig. \ref{fig:efield}(d-f), the charge density $\rho_e=e(n_i-n_e)$ around the rear-surface of the target is plotted, overlaid with the electric field lines (black), $\bold{E}=E_x\hat{x}+E_z\hat{z}$. As the magnetic vortex inside the channel exits the target,  it expands transversely and displaces the surface electrons. Therefore the rear-surface of the target is positively charged while the negative charge is concentrated on the apex of the filament with fast moving electrons along the central axis. The electric field lines emitting from the rear-surface converge onto the apex of the filament. Furthermore, the relativistic electrons  of the filament strengthen the transverse component of the electric field due to relativistic effects. Their velocity can be characterized by the group velocity of the laser pulse inside the channel, or in terms of $gamma$-factor, $\gamma_e=(\sqrt{2}/1.84)(2P/KP_c)^{1/6}(n_{cr}/n_e)^{1/3}$ \cite{bulanov2015}. The converging electric field indeed makes a contribution to the collimation of ions.

Figure \ref{fig:energy}(a) shows the transverse momentum $p_x-p_y$ distribution of the ions at several energy levels centered at $E=50,96,143,190$, and $236$ MeV (Run I-2). The distribution reveals highly collimated accelerated ions. 
The influence of laser power on the divergence of the ion beam is examined, as shown in Fig. \ref{fig:energy}(b) (Run I-1 to I-4); the divergence is defined as $2\Delta\theta= 2(\langle\theta^2\rangle-\langle\theta\rangle^2)$, where $\langle\,\,\rangle$ is the average of the angular distribution in each energy bin, $\theta=\text{cos}^{-1}\left[p_z/p\right]$, and $p_z$ is the $z$-component moment.
The energy is normalized by the maximum ion energy in each run.
The divergence $2\Delta\theta$ is around $5$ to $8$ degrees between $0.3<E/E_\text{max}<0.8$ for the laser powers, $0.35<P(\text{PW})<1.96$. 
All the runs reveal achromatic divergence. The converging electric field behind the rear surface of the target makes a contribution to such an achromatic divergence of the ion beam. 
Such a narrow angular dispersion is not commonly found in other acceleration schemes.
Exceptionally, a recent experiment on TNSA with a large laser focal diameter $2w_0\sim 100\lambda$ performed at the peta-watt BELLA laser facility revealed an achromatic divergence with $2\Delta\theta=100\text{mrad}=5.7^\circ$ of the ion beam \cite{steinke18}. In Fig. \ref{fig:efield}(f), the histogram shows the number of the accelerated ions in each energy bin with the size of $\Delta E=13$MeV, and 
about $5\times10^{10}$ ($2\times10^{10}$) ions are accelerated above $E=50$ ($100$) MeV.
We measure that $18\%$ (or $20\%$) of the laser energy is transferred to the total ions,
and $3.3\%$ (or $4.4\%)$ is transferred to the ions above $E=50$ MeV 
for P=1PW of Run I-2 (or P=2PW of Run I-1).
More than $50\%$ of the laser energy is used to heat the electrons for both cases.

\begin{figure}
\begin{center}
\includegraphics[scale=0.5]{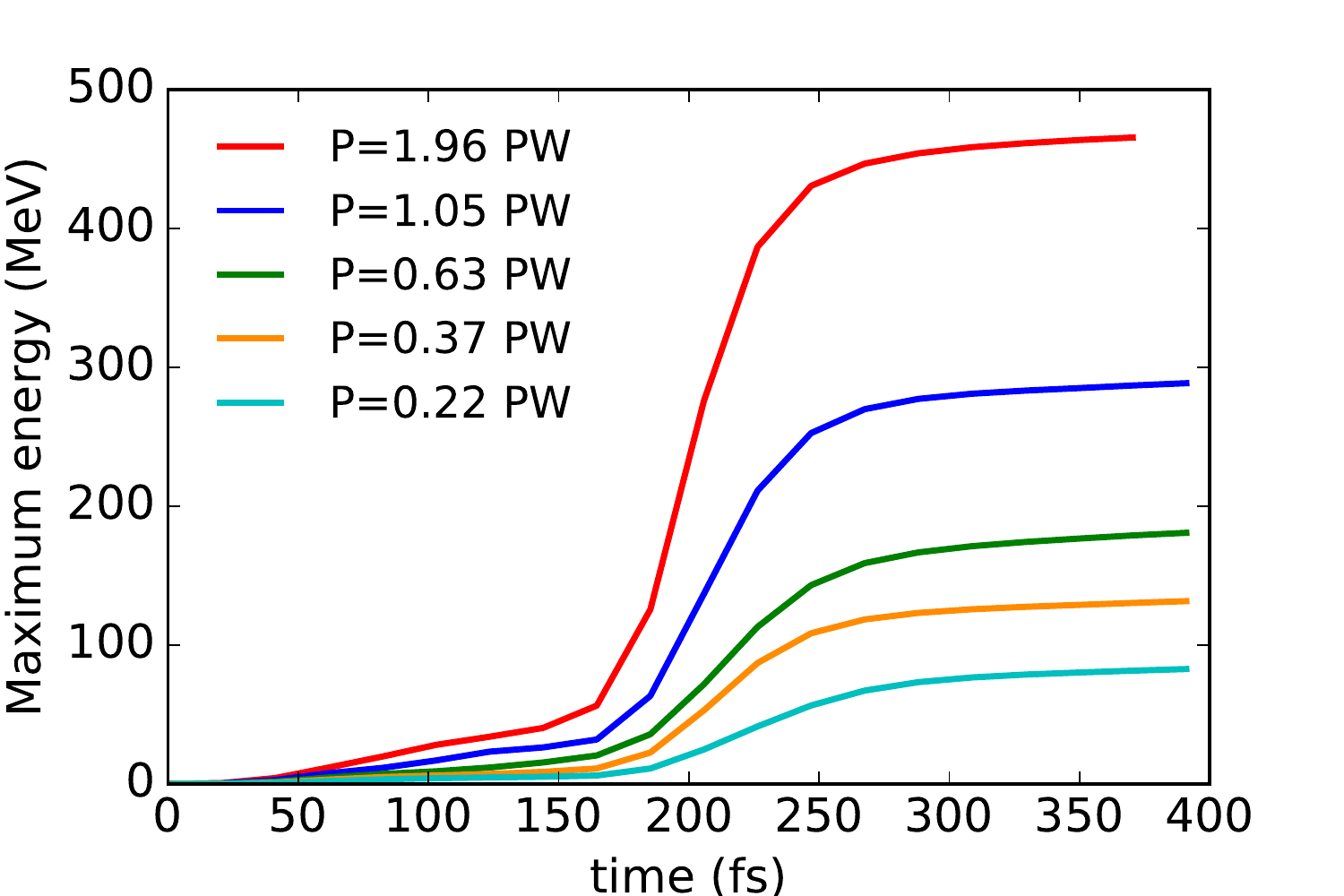}
\caption{Maximum ion energy as a function of time for different laser powers (Run I-1 to I-5). }
\label{fig:div_ene}
\end{center}
\end{figure}

Figure \ref{fig:div_ene} shows time evolution of the maximum ion energy for different laser powers (Run I-1 to I-4).
Here, the peak of the laser pulse arrives at the front surface at $t=53$fs.
The ions start to be accelerated around $t=180$fs after the laser pulse exits the target at $t=170$fs.
The ion gains about $80\%$ of the maximum energy during the acceleration stage from $t=180$ fs to $240$fs,
and reaches saturation at $t>300$ fs. 
The accelerated ions come from the edge of the filament 
when the channel in the ion distribution opens at the rear side of the target,
which lags behind the channel in the electron distribution.

\begin{figure}
\begin{center}
\includegraphics[scale=0.5]{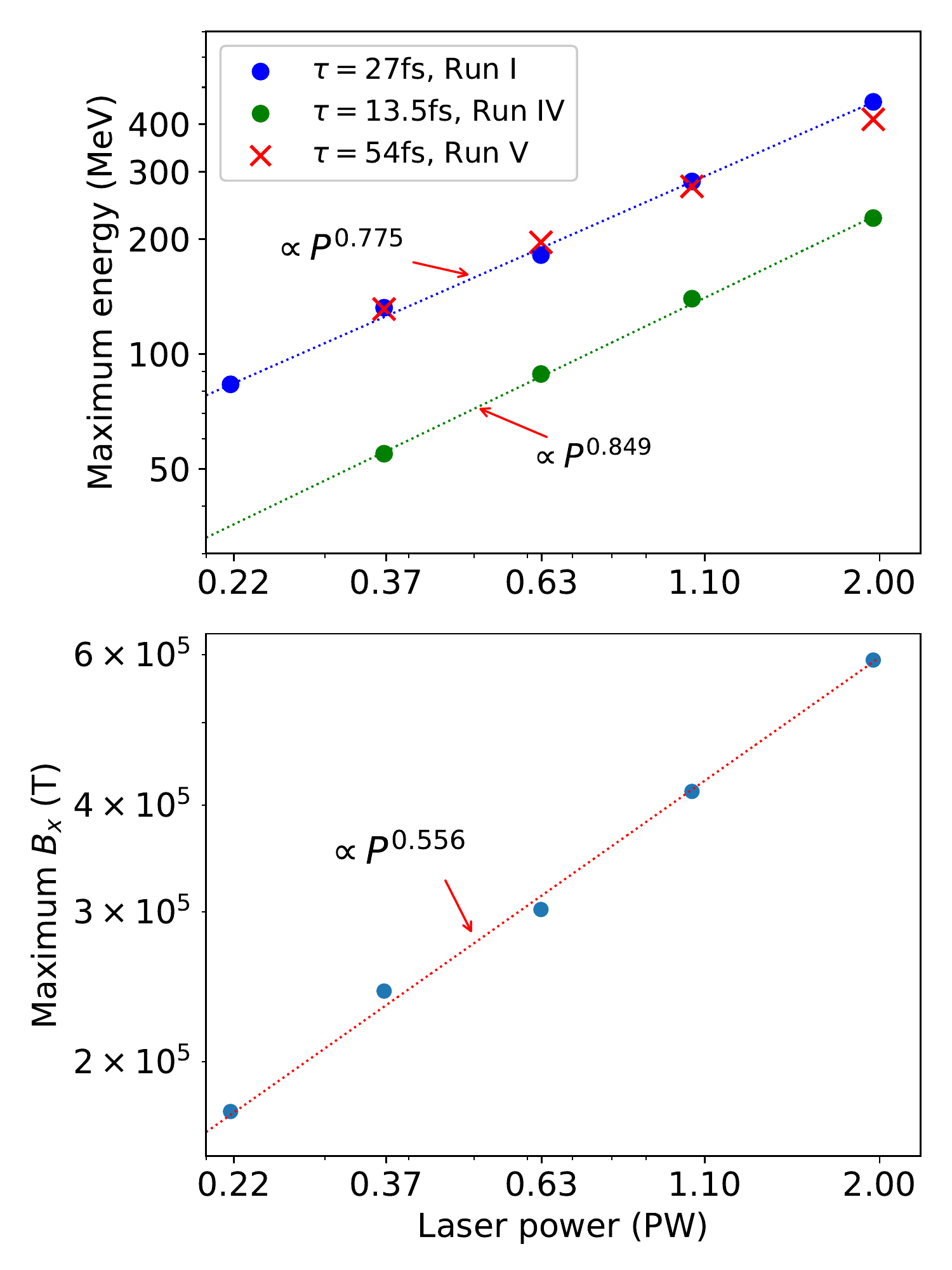}
\caption{(Top) Maximum ion energy vs. laser power for
$\tau=27$fs (Run I-1 to I-5), $\tau=13.5$fs (Run IV-1 to IV-4 ), and $\tau=54$fs (Run V-1 to V-4).
The power-law scaling is $\propto P^{0.775}$ for $\tau=27$fs (blue dashed) and 
$\propto P^{0.849}$ $\tau=13.5$fs (green dashed).
(Bottom) Maximum azimuthal $B$ field inside the channel vs. laser power 
for $\tau=27$fs (Run I-1 to I-5). The power-law scaling is $\propto P^{0.55}$.}
\label{fig:statistics}
\end{center}
\end{figure}

Figure \ref{fig:statistics} (top) shows the maximum ion energy vs. the laser power from 
Run I-1 to I-5 ($\tau=27$fs), IV-1 to IV-4 ($\tau=13.5$fs), and V-1 to V-4 ($\tau=54$fs) in Table \ref{setuptable}.
Interestingly, the data points fit to a power-law, $E_\text{max}\propto P^{\sigma}$, only when the spot size matches the channel radius, and the power-law index is $\sigma\sim0.8$. This scaling can be explained in the framework of a simple analytical model. We assume that the ions are accelerated by a pulsed longitudinal electric field $E$, which is $E=B_{max}$, and the length of this field is $R_\text{ch}$. We also assume that this field moves with the speed of light. Then for the ions, we can write an equation of motion: $dp/dt=\varepsilon(x-ct)$, where $\varepsilon=eE/m_i\omega c$ is the normalized electric field, time is normalized to $\omega^{-1}$, and the momentum is normalized to the ion mass times the speed of light. This equation can be solved in quadratures as $(-1+3p+p^3+(1+p^2)\sqrt{1+p^2})=6\pi R_\text{ch}/\lambda$. Here, we assumed that $\varepsilon$ is constant over the phase interval $(0,R_\text{ch})$, and is equal to zero elsewhere. This solution gives a scaling of $E_\text{max}\sim P$ for $P\ll 1$ PW, and for $P\sim 1$ PW the solution is well reproduced by the $P^{0.8}$ scaling, which is in good agreement with the results of 3D PIC simulations.

In Fig. \ref{fig:statistics} (top),
the maximum ion energy for the short pulse $\tau=13.5$fs (green dot) is reduced $50\%$ compared to that of
$\tau=27$fs (blue dot) for a given laser power
because the total laser energy is a half.
Interestingly, the maximum ion energy (red x) for the longest pulse duration $\tau=54$fs
is comparable to that of $\tau=27$fs.
Note that the parameters for the longest pulse do not meet the optimum condition in Eq.(\ref{opteq}).
To meet the optimum condition, either the density $n_e$ or the target
length $L_\text{ch}$ needs to be twice.
For $\tau=54$fs, the residual part of the laser pulse longer than $27$fs
escapes the target through the channel 
without heating the electrons inside, which is confirmed by the simulation (not shown in figures).
Therefore, the maximum ion energy for a given laser power 
is similar between the two pulses, $\tau=27$fs and $54$fs.
If the parameters for the pulse duration $\tau=54$fs meet the optimum condition in Eq.(\ref{opteq}), we predict that the maximum ion energy will increase significantly.

Figure \ref{fig:statistics} (bottom) shows the maximum magnetic field generated inside the channel
vs. the laser power from Run I-1 to I-5 ($\tau=27$fs).
The maximum field strength is localized inside the leading edge of the channel,
$B\sim 0.4$MT (or $0.6$MT) for the laser power $P=1$PW (or $2$PW).
The power-law scaling $B_\text{max}\propto P^{0.55}$ is close to the theoretical prediction
where the power-law index is given by $1/2$ \cite{bulanov2015}.

In this section, the ion beams revealed high level of collimation and  achromatic angular divergence due to the converging electric field behind the target. Such a significant amount of charge $\sim10^{10} H^{+}$ above $100$MeV
for $P\sim 1$PW makes the MVA scheme potentially a desirable source for the application to hadron therapy.
A favorable energy-power scaling $E_\text{max}\propto P^{0.8}$ is obtained from the 3D PIC simulations, and is also predicted from our simple analytic models as long as the optimum conditions, Eq.(\ref{rcheq}) and (\ref{opteq}), are satisfied.

\section{Summary and conclusion}

We studied laser driven ion acceleration in the MVA regime by employing 3D PIC simulations and analytic models. In order to optimize the process of acceleration from the point of view of increasing maximum ion energy we studied the coupling of the laser pulse to the target by a comprehensive parameter scan. We varied the focal spot size, the power, and the duration of the laser, as well as the density of the target.

We showed that the optimal acceleration happens when two conditions are satisfied. First, the laser is focused at the front of the target to a spot, which radius is equal to the radius of the laser-generated channel in the target, and, second, the density and thickness of the target are given by the laser depletion condition. 

The 3D computer simulations revealed the structure of the electric and magnetic fields inside the target, as well as that of the plasma itself. It was shown that the laser creates a co-axial plasma structure in the target along the direction of its propagation: a laser-generated channel with high density wall and a strong pinched electron current, flowing along the channel central axis. A strong magnetic field is created by the electric current flowing in the filament and the return current, flowing in the wall. The co-axial structure of the currents ensures that the magnetic field is localized inside the channel. Due to the strong pinching of the central filament, which is due to the fact that the electrons have relativistic energies, the magnetic field amplitude can almost reach MT-level ($6\times 10^5$ Tesla observed in a simulation for Run I-1, a 2 PW case), which is in good agreement with analytical estimates \cite{bulanov2015}.

We showed that as the intense laser-driven electron current and magnetic field leave the target from its back, strong longitudinal and transverse electric fields are established. These fields accelerate the protons to several hundred MeV maximum energy and collimate them into  a well defined beam with achromatic angular divergence with $2\Delta \theta \sim 7^o$.

The 3D PIC simulation results prove the validity of the waveguide analytical model of the MVA regime proposed in Refs \cite{bulanov2010,bulanov2015}. Using this model we were able to analyze the scaling of the magnetic field in the channel, which scales as the square root of laser power, and the scaling of the maximum ion energy, which scales as (laser power)$^{0.8}$ for laser pulse power ranging from 0.2 PW to 2 PW. We note that the scaling works only for the optimal coupling of the laser to the target. 

\begin{acknowledgments}
We especially thank the WarpX developer team in LBNL, LLNL, and SLAC for invaluable help.
Warpx has been newly developed for plasma accelerator modeling under the U.S. DoE's Exascale Computing Project. 
J. Park thanks Maxence Th\'evenet and R\'emi Lehe for helpful assistance.
J. H. Bin acknowledges financial support from the Alexander von Humboldt Foundation. We gratefully acknowledge the discussions with D. Margarone and G. Korn.
This work was supported by LDRD funding from LBNL provided by the Director, and the U.S. DOE Office of Science Offices of HEP and FES, under Contract No. DE-AC02-05CH11231. This research used computational resources (Cori) of the National Energy Research Scientific Computing center (NERSC), which is supported by the Office of Science of the U.S. DOE under Contract No. DE-AC02-05CH11231. An award of computer time was provided by the INCITE program. This research used resources of the Argonne Leadership Computing Facility, which is a DOE Office of Science User Facility supported under Contract DE-AC02-06CH11357.
This research was supported in part by the Exascale Computing Project (ECP), Project
Number: 17-SC-20-SC, a collaborative effort of two DOE organizations---the
Office of Science and the National Nuclear Security
Administration---responsible for the planning and preparation of a capable
exascale ecosystem---including software, applications, hardware, advanced
system engineering, and early testbed platforms---to support the nation's
exascale computing imperative.

\end{acknowledgments}


\begin{thebibliography}{99}

\bibitem{MTB_review} G. Mourou, T. Tajima, and S.  V. Bulanov, Rev. Mod. Phys. {\bf 78}, 309 (2006).

\bibitem{Daido_review} H. Daido, M. Nishiuchi, and A. S. Pirozhkov, Reports on Progress in Physics {\bf 75}, 056401 (2012).

\bibitem{Macchi_review} A. Macchi, M. Borghesi, and M. Passoni, Rev. Mod. Phys. {\bf 85}, 751 (2013).

\bibitem{Bulanov_review} S. V. Bulanov, Ja. J. Wilkens, T. Zh. Esirkepov, G. Korn, G. Kraft, S. Kraft, M. Molls, 
and V. S. Khoroshkov, Phys. Usp. \textbf{57}, 1149 (2014).

\bibitem{BELLA} K. Nakamura, H. S. Mao, A. J. Gonsalves, H. Vincenti, D. E. Mittelberger, J. Daniels, A. Magana, C. Toth, and W. P. Leemans, IEEE Journal of Quantum Electronics \textbf{53}, 1 (2017).

\bibitem{ELI-NP} S. Gales, K. A. Tanaka, D. L. Balabanski, F. Negoita, D. Stutman, O. Tesileanu, C. A. Ur, D. Ursescu, I. Andrei, S. Ataman, M. O. Cernaianu, L. DAlessi, I. Dancus, B. Diaconescu, N. Djourelov, D. Filipescu, P. Ghenuche, D. G. Ghita, C. Matei, K. Seto, M. Zeng, and N. V. Zamfir, Rep. Prog. Phys. \textbf{81}, 094301 (2018).

\bibitem{ELI-BL} ELI-beamlines, https://www.eli-beams.eu/.

\bibitem{CORELS} CORELS, https://www.ibs.re.kr/eng/sub02\_03\_05.do.

\bibitem{XCELS} XCELS, http://www.xcels.iapras.ru

\bibitem{VULCAN} VULCAN, https://www.clf.stfc.ac.uk/Pages/Vulcan-2020.aspx.

\bibitem{Danson_2015} C. Danson, D. Hillier, N. Hopps, and D. Neely, High Power Laser Science and Engineering \textbf{3}, e3 (2015).

\bibitem{RPA_93 MeV} I. J. Kim, K. H. Pae, C. M. Kim, C.-L. Lee, I. W. Choi, H. T. Kim, H. Singhal, J. H. Sung, S. K. Lee, H. W. Lee, P. V. Nickles, T. M. Jeong, and C. H. Nam, Phys. Plasmas \textbf{23}, 070701 (2016).

\bibitem{TNSA_85 MeV} F. Wagner, O. Deppert, C. Brabetz, P. Fiala, A. KleinSchmidt, P. Poth, V. A. Schanz, A. Tebartz, B. Zielbauer, M. Roth, et al., Phys. Rev. Lett. \textbf{116}, 205002 324 (2016).

\bibitem{TNSA_RPA_94 MeV} A. Higginson, R. J. Gray, M. King, R. J. Dance, S. D. R. Williamson, N. M. H. Butler, R. Wilson, R. Capdessus, C. Armstrong, J. S. Green, et al., Nature Comm. \textbf{9}, 724 (2018).


\bibitem{snavely2000} R. A. Snavely, M. H. Key, S. P. Hatchett, T. E. Cowan, M. Roth, T. W. Phillips, M. A. Stoyer, E. A. Henry, T. C. Sangster, M. S. Singh, S. C. Wilks, A. MacKinnon, A. Offenberger, D. M. Pennington, K. Yasuike, A. B. Langdon, B. F. Lasinski, J. Johnson, M. D. Perry, and E. M. Campbell, Phys. Rev. Lett. {\bf85}, 14 (2000)

\bibitem{maksimchuck2000} A. Maksimchuk, S. Gu, K. Flippo, and D. Umstadter, and V. Yu. Bychenkov, Phys. Rev. Lett. {\bf84}, 18 (2000)

\bibitem{Wilks_2001} S. C. Wilks, A. B. Langdon, T. E. Cowan, M. Roth, M. Singh, S. Hatchett, M. H. Key, D. Pennington, A. MacKinnon, and R. A. Snavely, Phys. Plasmas \textbf{8}, 542 (2001).

\bibitem{Fuchs_2006} J. Fuchs, P. Antici, E. D'Humieres, E. Lefebvre, M. Borghesi, E. Brambrink, C. A. Cecchetti, M. Kaluza, V. Malka, and M. Manclossi, Nature Physics \textbf{2}, 48 (2006)

\bibitem{RPA} T. Esirkepov, M. Borghesi, S. V. Bulanov, G. Mourou, and T. Tajima, Phys. Rev. Lett. {\bf 92}, 175003 (2004).

\bibitem{henig2009} A. Henig, S. Steinke, M. Schnu\"rer, T. Sokollik,3 R. Ho\"rlein, D. Kiefer, D. Jung,
J. Schreiber, B. M. Hegelich, X. Q. Yan, J. Meyer-ter-Vehn, T. Tajima,
P. V. Nickles, W. Sandner, and D. Habs, Phys. Rev. Lett. {\bf 103}, 245003 (2009)

\bibitem{bin2015} J. H. Bin, W. J. Ma, H. Y. Wang, M. J. V. Streeter, C. Kreuzer, D. Kiefer,
M. Yeung, S. Cousens, P. S. Foster, B. Dromey, X. Q. Yan,3 R. Ramis,
J. Meyer-ter-Vehn, M. Zepf, and J. Schreiber, Phys. Rev. Lett. {\bf 115}, 064801 (2015)

\bibitem{SWA} D. Haberberger, S. Tochitsky, F. Fiuza, C. Gong, R. A. Fonseca, L. O. Silva, W. B. Mori, and C. Joshi, Nature Physics {\bf 8}, 95 (2012)

\bibitem{RIT} S. Palaniyappan, B. M. Hegelich, H.-C. Wu, D. Jung, D. C. Gautier, L. Yin, B. J. Albright, R. P. Johnson, 
T. Shimada, S. Letzring, D. T. Offermann, J. Ren, C. Huang, R. Horlein, B. Dromey, J. C. Fernandez, and R. C. Shah, 
Nature Physics {\bf 8}, 763 (2012).

\bibitem{MVA} A. V. Kuznetsov, T. Zh. Esirkepov, F. F. Kamenets, and S. V. Bulanov, Plasma Phys. Rep. {\bf 27}, 211 (2001); S. V. Bulanov and T. Zh. Esirkepov, Phys. Rev. Lett. {\bf 98}, 049503 (2007).

\bibitem{SSB_PoP} S. S. Bulanov,E. Esarey,C. B. Schroeder, S. V. Bulanov, T. Zh. Esirkepov, M. Kando, F. Pegoraro, W. P. Leemans, Phys. Plasmas \textbf{23}, 056703 (2016).

\bibitem{willingale2006} L. Willingale, S. P. D. Mangles, P. M. Nilson, R. J. Clarke, A. E. Dangor, M. C. Kaluza, S. Karsch, K. L. Lancaster, W. B. Mori, Z. Najmudin, J. Schreiber, A. G. R. Thomas, M. S. Wei, and K. Krushelnick, Phys. Rev. Lett. \textbf{96}, 245002 (2006).

\bibitem{fukuda2009} Y. Fukuda, A. Ya. Faenov, M, Tampo, T. A. Pikuz, T. Nakamura, H. Kando, Y. Hayashi, A. Yogo, H. Sakaki, T. Kameshima, A. S. Pirozhkov, K. Ogura, M. Mori, T. Zh. Esirkepov, J. Koga, A. S. Boldarev, V. A. Gasilov, A. I. Magunov, T. Yamauchi, R. Kodama, P. R. Bolton, Y. Kato, T. Tajima, H. Daido, and S. V. Bulanov, Phys. Rev. Lett. \textbf{103}, 165002 (2009).

\bibitem{willingale2011} L. Willingale, P. M. Nilson, A. G. R. Thomas, S. S. Bulanov, A. Maksimchuk, W. Nazarov, T. C. Sangster, C. Stoeck, and K. Krushelnick, Phys. Plasmas \textbf{18}, 056706 (2011).

\bibitem{helle2016} M. H. Helle, D. F. Gordon, D. Kaganovich, Y. Chen, J. P. Palastro, and A. Ting, Phys. Rev. Lett. \textbf{117}, 165001 (2016).

\bibitem{bulanov2010}  S. S. Bulanov, V. Y. Bychenkov, V. Chvykov, G. Kalinchenko, D. William L., 
T. Matsuoka, A. G. R. Thomas, L. Willingale, V. Yanovsky, K. Krushelnick, and A. Maksimchuk 
Phys of Plasmas {\bf17}, 043105 (2010) 

\bibitem{nakamura2010} T. Nakamura, S. V. Bulanov, T. Z. Esirkepov, and M. Kando  PRL {\bf105} 135002 (2010)

\bibitem{bulanov2015}  S. S. Bulanov, E. Esarey, C. B. Schroeder, W. P. Leemans, S. V. Bulanov, D. Margarone, G. Korn, and T. Haberer, Phys. Rev. Special Topics - Accelerators and Beams  {\bf18}, 061302 (2015)

\bibitem{sharma2016} A. Sharma and A. Andreev, Laser and Particle Beams  {\bf34} 219 (2016)

\bibitem{sharma2018} A. Sharma, Scientific Reports {\bf8} 2191 (2018)

\bibitem{bulanov.prl.2015} S.S. Bulanov, E. Esarey, C. B. Schroeder, S. V. Bulanov, T. Zh. Esirkepov, M. Kando, F. Pegoraro, and W. P. Leemans, Phys. Rev. Lett. 114, 105003 (2015).

\bibitem{brady.ppcf.2013} C. S. Brady, C. P. Ridgers, T. D. Arber, and A. R. Bell, Plasma Phys. Control. Fusion 55, 124016 (2013). 

\bibitem{zhu.natcomm.2016} X. L. Zhu, T. P. Yu, Z. M. Sheng, Y. Yin, I. C. E. Turcu, and A. Pukhov, Nature Communications 7, 13686 (2016).

\bibitem{liu.pop.2018} J.-x. Liu, Y. Zhao, X.-p. Wang, J.-z. Quan, T.-p. Yu, G.-B. Zhang, X.-h. Yang, Y.-y. Ma, F.-q. Shao, and J. Zhao, Physics of Plasmas 25, 103106 (2018).

\bibitem{gu.commphys.2018} Y.-J. Gu, O. Klimo, S. V. Bulanov, and S. Weber, Communications Physics 1, 93 (2018)

\bibitem{chen.ieee.1987} P. Chen, J. J. Su, T. Katsouleas, S. Wilks, and J. M.
Dawson, IEEE Trans. Plasma Sci. 15, 218 (1987).

\bibitem{sun1987} Guo-Zheng Sun, Edward Ott, Y. C. Lee, and Parvez Guzdar, Phys. Fluids  {\bf30} 2 (1987)
 
\bibitem{stark2016} D. J. Stark, T. Toncian, and A. V. Arefiev, Phys. Rev. Lett. 116, 185003 (2016);

\bibitem{jansen2018} O. Jansen, T. Wang, D. Stark, E. d’Humieres, T. Toncian, and A. Arefiev, Plasma Phys. Cointrol. Fusion 60, 054006 (2018)

\bibitem{arefiev2018} A. V. Arefiev, Z. Gong, T. Toncian, and S. S. Bulanov, arXiv:1807.07629 

\bibitem{vay2018}
J.-L. Vay, A. Almgren, J. Bell, L. Ge, D.P. Grote, M. Hogan, O. Kononenko, R. Lehe, A. Myers, C. Ng, 
J. Park, R. Ryne, O. Shapoval, M. Thvenet, and W. Zhang, 
Nuclear Inst. and Methods in Physics Research, A 01.035 (2018)

\bibitem{shapoval18} O. Shapoval, J.-L. Vay, H. Vincenti, Comp. Phys. Comm. (2018)

\bibitem{steinke18} S. Steinke, J. H. Bin, J. Park, Q. Ji, K. Nakamura, A. J. Gonsalves, S. S. Bulanov,
C. Toth, J-L. Vay, C. B. Schroeder, E. Esarey, T. Schenkel, and W. P. Leemans, submitted (2018)



\end{thebibliography}
\end{document}